\documentclass[journal,twoside,web]{ieeecolor}
\usepackage{tmi}
\usepackage{cite}
\usepackage{amsmath,amssymb,amsfonts}
\usepackage{algorithmic}
\usepackage{graphicx}
\usepackage{textcomp}
\usepackage{url}
\usepackage{multirow}
\usepackage{makecell} % 表格内换行

\def\BibTeX{{\rm B\kern-.05em{\sc i\kern-.025em b}\kern-.08em
    T\kern-.1667em\lower.7ex\hbox{E}\kern-.125emX}}
\begin{document}
\title{BAAF: A benchmark attention adaptive framework for medical ultrasound image segmentation tasks}
% \author{First A. Author, \IEEEmembership{Fellow, IEEE}, Second B. Author,and Third C. Author, Jr., \IEEEmembership{Member, IEEE}
\author{Gongping Chen, Lei Zhao, Xiaotao Yin, Liang Cui, Jianxun Zhang, Yu Dai, Ningning Liu 
\thanks{This work is supported by the National Natural Science Foundation of China (U1913207) and the Tianjin Research Innovation Project for Postgraduate Students (2022BKY004). }
	%   (Corresponding author: Yu Dai, e-mail: daiyu@nankai.edu.cn).
\thanks{Gongping Chen with the College of Biomedical Engineering and Technology, Tianjin Medical University, Tianjin, China. (Corresponding author, e-mail: cgp110@mail.nankai.edu.cn). }
%\thanks{Rui Wang with theDepartment of Orthopedic Surgery, Tianjin Medical University General Hospital, Tianjin, China (e-mail: drwangrui@outlook.com).}
\thanks{Lei Zhao with the College of Computer Science and Electronic Engineering, Hunan University, Changsha, China.}
\thanks{Xiaotao Yin with the Department of Urology, Fourth Medical Center of PLA General Hospital, Beijing, China.}
\thanks{Liang Cui with the Department of Urology, Civil Aviation General Hospital, Beijing, China.}
%\thanks{Yan Wang, with the Department of Breast Cancer, Tianjin Medical University Cancer Institute and Hospital, National Clinical Research Centre for Cancer, Tianjin, China (e-mail: Dr.wyan@outlook.com).}
% \thanks{Jianxun Zhang with the College of Artificial Intelligence, Nankai University, Tianjin, China. }
\thanks{Jianxun Zhang, Yu Dai with the College of Artificial Intelligence, Nankai University, Tianjin, China. }
\thanks{Ningning Liu with the Department of Ultrasonography, Tianjin Medical University Second Hospital, Tianjin, China. }}
\maketitle

\begin{abstract}
The AI-based assisted diagnosis programs have been widely investigated on medical ultrasound images. Complex scenario of ultrasound image, in which the coupled interference of internal and external factors is severe, brings a unique challenge for localize the object region automatically and precisely in ultrasound images. In this study, we seek to propose a more general and robust Basic Attention Adaptive Framework (BAAF) to assist doctors segment or diagnose lesions and tissues in ultrasound images more quickly and accurately.The BAAF is an optimisation of existing attention schemes, which consists of a parallel hybrid attention module (PHAM) and an adaptive calibration mechanism (ACM). Specifically, BAAF first coarsely calibrates the input features from the channel and spatial dimensions, and then adaptively selects more robust lesion or tissue characterizations from the coarse-calibrated feature maps. The design of BAAF further optimizes the “what” and “where” focus and selection problems in CNNs and seeks to improve the segmentation accuracy of lesions or tissues in medical ultrasound images. The method is evaluated on four medical ultrasound segmentation tasks, and the adequate experimental results demonstrate the remarkable performance improvement over existing state-of-the-art methods. The Dice values of the method on the two breast datasets are 79.05$\%$ and 80.85$\%$, and on the two kidney datasets are 94.20$\%$ and 91.29$\%$. The two kidney ultrasound datasets will be released with the paper. This work provides the possibility for automated medical ultrasound assisted diagnosis. 

%%The source code is publicly available on \textcolor{magenta}{\underline {https://github.com/CGPxy/BAAF}}.
\end{abstract}


%% 文件目录
\graphicspath{{Fig1/}} 
%% h：当前位置，t：顶端，b：下，p：浮动 ；ht，htbp组合
\begin{figure}[ht]\footnotesize
	\centerline{\includegraphics[scale=.48]{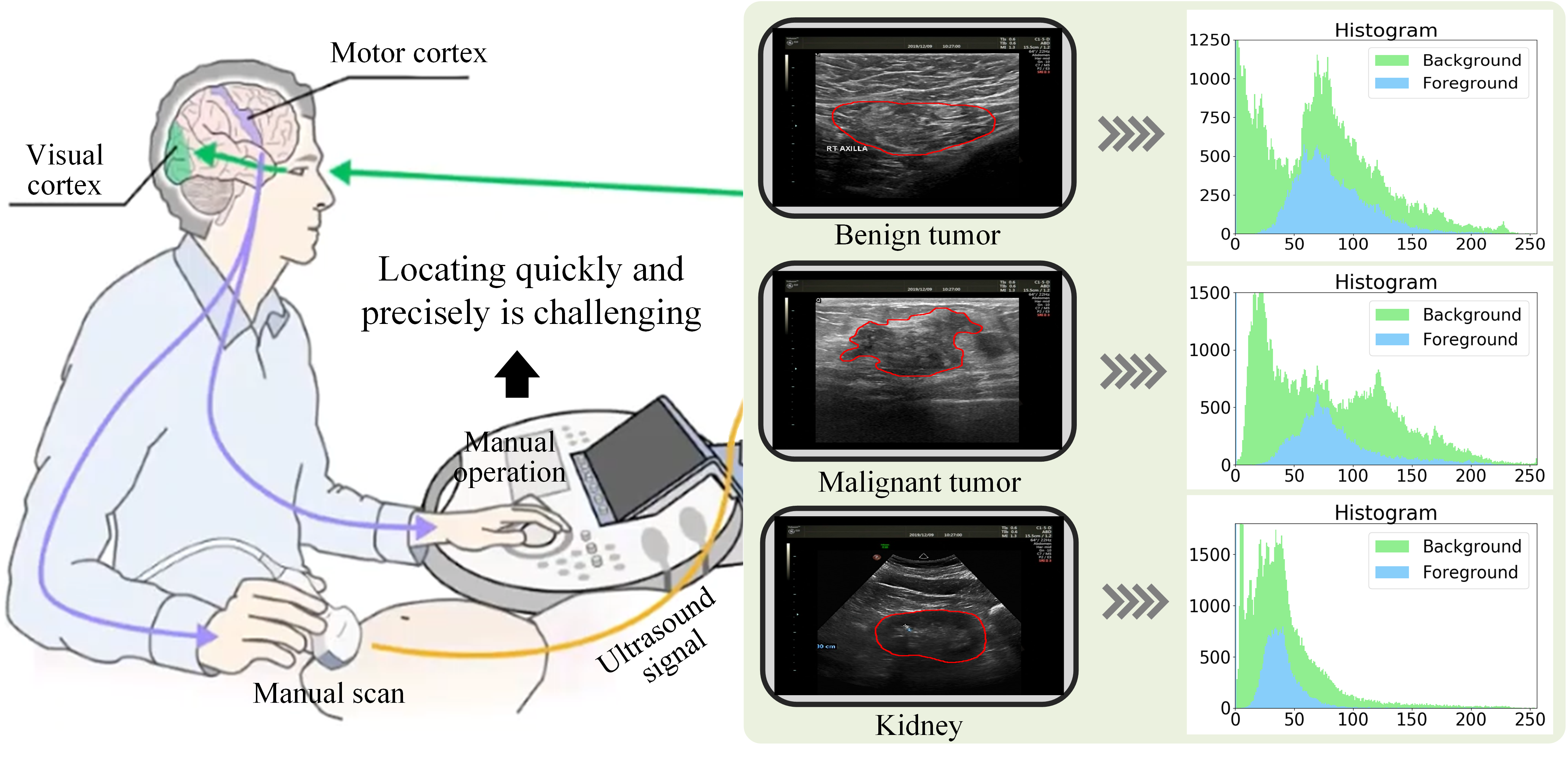}}
	\begin{sloppypar}
		\textbf{Fig. 1.} Manual annotation is a pain point due to the perturbation of similar intensity distributions, variable subject morphology, and other factors.
	\end{sloppypar}
\end{figure}

\section{Introduction}
\label{sec:introduction}
\IEEEPARstart{C}{ancer} has long been a major public health problem worldwide. “Cancer statistics, 2023” shows that the average number of deaths due to cancer per day is 1,670, and it also suggests that early clinical screening remains one of the most important means of improving survival \textcolor{blue}{\cite{Siegel2023}}. Ultrasound imaging, as one of the most common imaging schemes in clinical practices, has been extensively applied to the early detection of many diseases (e.g., breast cancer and kidney) in view of its safety and efficiency \textcolor{blue}{\cite{Noble2006}}. Clinically, due to the lack of a gold standard for diagnosis, the accuracy of diagnosis largely depends on the skill level of radiologists, which is experience-dependent and may suffer from high inter-observer variation even for well-trained radiologists \textcolor{blue}{\cite{Chen2023}}. In recent years, many AI-based computer-aided diagnostic (CAD) systems for medical ultrasound have been developed to help radiologists improve diagnostic reliability and reduce individual subjectivity, in which lesion and tissue segmentation is one of the key steps \textcolor{blue}{\cite{Liu2019}}.

 As shown in Fig. \textcolor{red}{1}, complex scenario of ultrasound image, in which the coupled interference of internal and external factors is severe, brings a unique challenge for radiologists to identify lesions and tissues quickly and accurately in ultrasound images. Therefore, the study of automatic segmentation methods for medical ultrasound images is crucial to improve efficiency and reduce misdiagnosis. Following the burgeoning of AI technology, many deep learning algorithms based on U-net and FCNN have been developed to achieve the automatic segmentation of medical ultrasound images \textcolor{blue}{\cite{Ronneberger2015, Long2015, Wang2020}}. Wu et al. employed the cascaded FCNN framework to refine the segmentation results of prenatal ultrasound images step-by-step \textcolor{blue}{\cite{Wu2017}}. Kim et al. seek to achieve the automatic segmentation of coronary arteries in ultrasound images by introducing multi-scale inputs and hybrid loss functions in U-net \textcolor{blue}{\cite{Kim2018}}. Abraham et al., designed a novel focal tversky loss to improve the segmentation performance of the deep learning model (MADU-net) in breast ultrasound images \textcolor{blue}{\cite{Abraham2019}}. Subsequently, Yap et al., explored the performance of U-net and FCNN on the segmentation task of breast ultrasound images \textcolor{blue}{\cite{Yap2018}}. Although these CNN frameworks significantly improves the diagnostic efficiency of medical ultrasound images, similar intensity distributions, variable lesion morphology, and other factors still constrain the possibility of implementing these AI models in the clinical practice.
 
 %% 文件目录
 \graphicspath{{Fig1/}} 
 %% h：当前位置，t：顶端，b：下，p：浮动 ；ht，htbp组合
 \begin{figure}[!t]\footnotesize
 	\includegraphics[scale=.57]{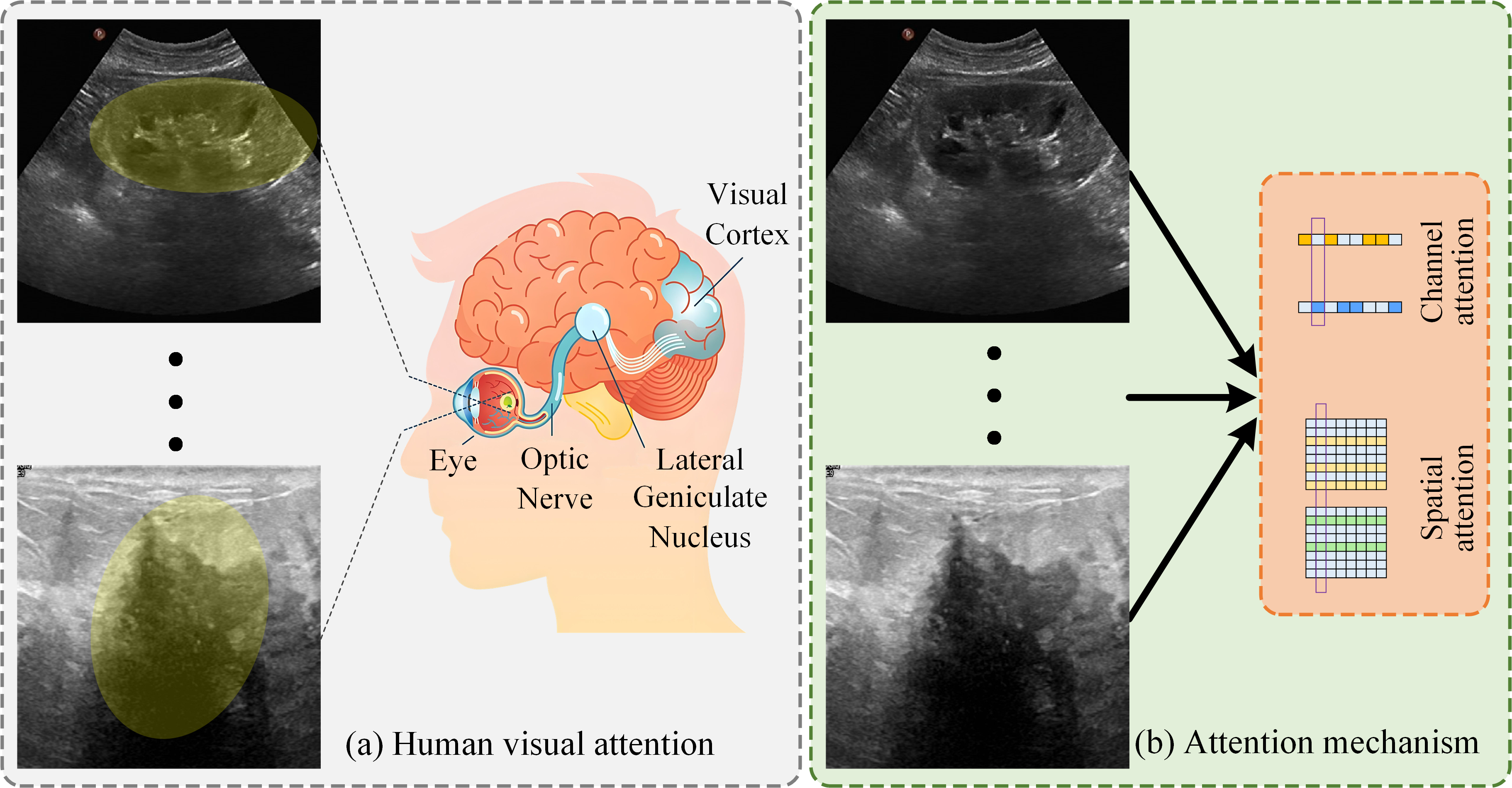}
 	\centering
 	\begin{sloppypar}
 		\textbf{Fig. 2.} The schematic of human visual attention and typical attentions.
 	\end{sloppypar}
 \end{figure}
 
To further improve the performance of CNNs in ultrasound segmentation tasks, there have been many attempts to integrate attention mechanism into CNNs \textcolor{blue}{\cite{Xian2018}}. It is well known that attention plays a very important role in human perception, and the introduction of the attention module can help the network to better focus on the “what” and “where” of the object from the whole image \textcolor{blue}{\cite{Ru2023}}, as shown in Fig. \textcolor{red}{2}. Zhuang et al. designed a Residual-Dilated-Attention-Gate-UNet (RDAU-Net) based on a spatial attention model to achieve breast tumor ultrasound segmentation \textcolor{blue}{\cite{Zhuang2019}}. Lee et al. attempted to segment breast ultrasound images by introducing a channel attention module with multi-scale average pooling operation in U-Net \textcolor{blue}{\cite{Lee2020}}. Xian et al., constructed a attention enhancement module (AENet) by using saliency maps to improve the segmentation accuracy of breast lesions \textcolor{blue}{\cite{Vakanski2020}}. Yan et al., proposed an attention enhanced U-net with hybrid dilated convolution (AE U-net) model to segment the breast ultrasound image \textcolor{blue}{\cite{Yan2022}}. Chen et al., developed a hybrid attention model with the spatial attention and the channel attention to segment kidney ultrasound images \textcolor{blue}{\cite{Chen2022}}. Meng et al., proposed a novel dual global attention neural network (DGANet) by integrating a bilateral spatial attention module and a global channel attention module to improve the detection accuracy of breast lesion \textcolor{blue}{\cite{Meng2023}}. However, it is still difficult for the individual attention (e.g., channel attention and spatial attention) or the simple hybrid attention mechanisms to localize and capture the objective characterization accurately from the complex ultrasound images, as shown in Fig. \textcolor{red}{3}.

To further mitigate the impact of variable objective morphology on segmentation accuracy, many works have attempted to combine the multi-scale feature information with the attention mechanism. Lyu et al., designed a pyramid attention network combining attention mechanism and multi-scale features (AMS-PAN) to improve the characterization of breast tumors in ultrasound images \textcolor{blue}{\cite{Lyu2023}}. Byra et al., constructed a novel CNN (SKNet) by integrating selective kernel (SK) convolution, which can help the network to select features that better characterize breast tumors under different receptive fields \textcolor{blue}{\cite{Byra2020}}. Similarly, Chen et al., proposed AAU-net \textcolor{blue}{\cite{Chen2023}} and ESKNet \textcolor{blue}{\cite{Chen2022b}} successively by introducing spatial dimension calibration in the selective kernel (SK) convolution, and they were utilized on medical ultrasound image segmentation tasks. Notably, these studies are more focused on extracting robust characteristic information under multi-scale receptive fields. Although being under the multi-scale receptive field can help the network to extract more robust feature representations, it remains subject to the influence of low-level or non-correlated features.
%% 文件目录
\graphicspath{{Fig1/}} 
%% h：当前位置，t：顶端，b：下，p：浮动 ；ht，htbp组合
\begin{figure}[!t]\footnotesize
	\includegraphics[scale=.47]{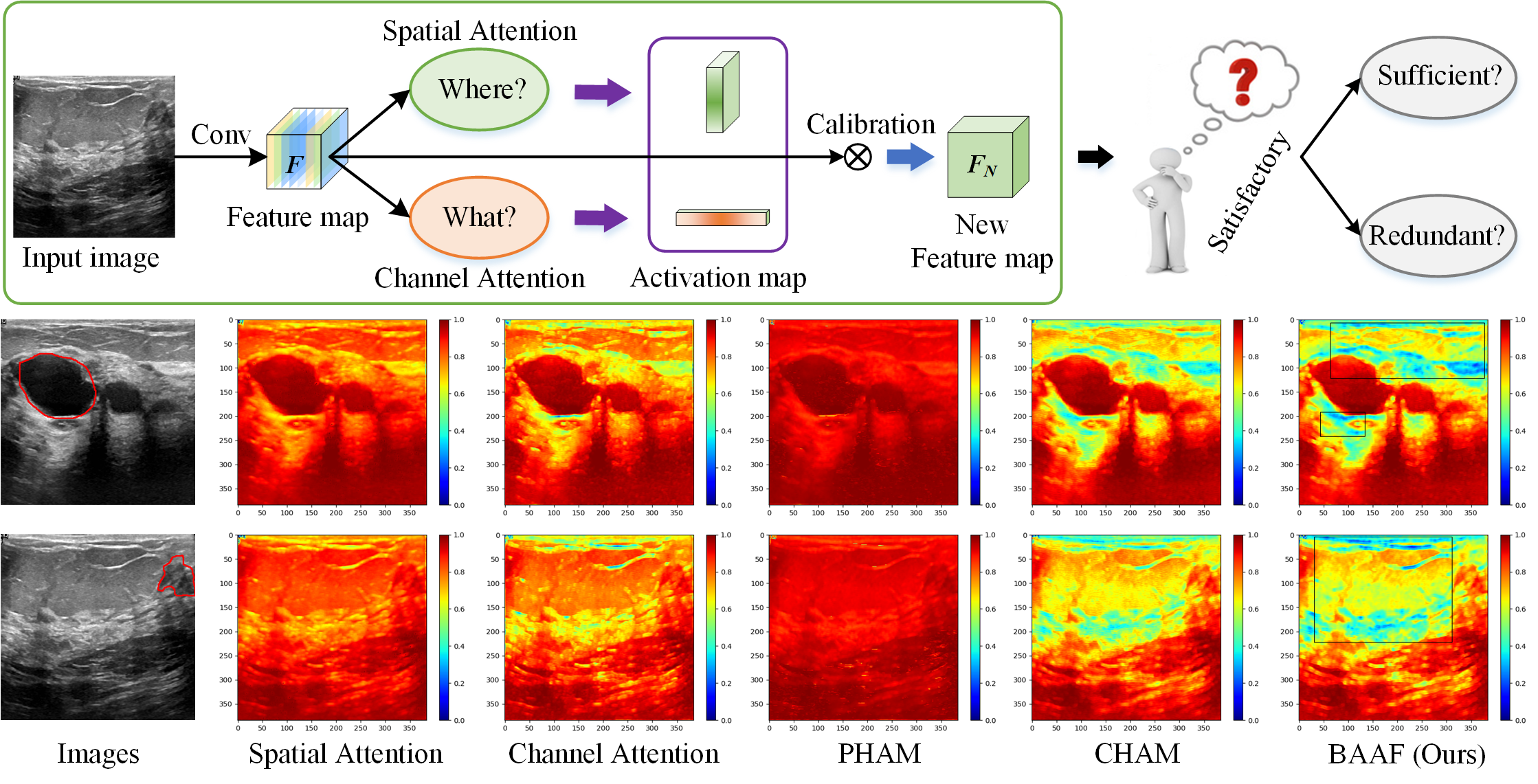}
	%\centering
	\begin{sloppypar}
		\textbf{Fig. 3.} The heat map of different attention mechanisms. The heat map shows the limitations of the single attention or simple combinations of the individual attention, with much false detection and missed detection. The detailed descriptions of spatial attention, channel attention, parallel hybrid attention mechanism (PHAM), cascade hybrid attention mechanism (CHAM) and basic attention adative framework (BAAF) can be seen in Fig. 4.
	\end{sloppypar}
\end{figure}

To cope with the above challenges, the adaptive operation on the initial feature maps calibrated by the attention mechanism will bring more benefit. In this work, we designed a more general and robust Basic Attention Adaptive Framework (BAAF) to further improve the ability of CNNs to characterize medical ultrasound images. As shown in Fig. \textcolor{red}{4}, Comparing to existing mechanisms of spatial attention, channel attention and their typical combinations, BAAF can adaptively select the feature information calibrated by the channel attention module and the spatial attention module. Specially, the BAAF module mainly contains a parallel hybrid attention mechanism (PHAM) and an adaptive calibration mechanism (ACM). In general, the main contributes as follows:

\begin{itemize}
	\item First, we proposed a Basic Attention Adaptive Framework (BAAF), which not only helps the network to locate the “what” and “where” that it should focus on more quickly, but also can adaptively select the calibrated “what” and “where”.
	\item Second, extensive experimental results on four public datasets demonstrate that the introduction of BAAF can further improve the segmentation accuracy of lesions or tissues in medical ultrasound images, which provides the possibility of clinically assisted diagnosis.
	% \item Third, BAAF still shows satisfactory segmentation ability during the comparison with existing attention method.
	\item Moreover, we contribute two clinical datasets to the medical ultrasound community, which are the kidney ultrasound dataset (KUS) and the kidney-cyst ultrasound dataset (KCUS). They will be available with the paper.
\end{itemize}

%% 文件目录
\graphicspath{{Fig1/}} 
%% h：当前位置，t：顶端，b：下，p：浮动 ；ht，htbp组合
\begin{figure*}[!t]\footnotesize
	\centering
	\includegraphics[scale=0.8]{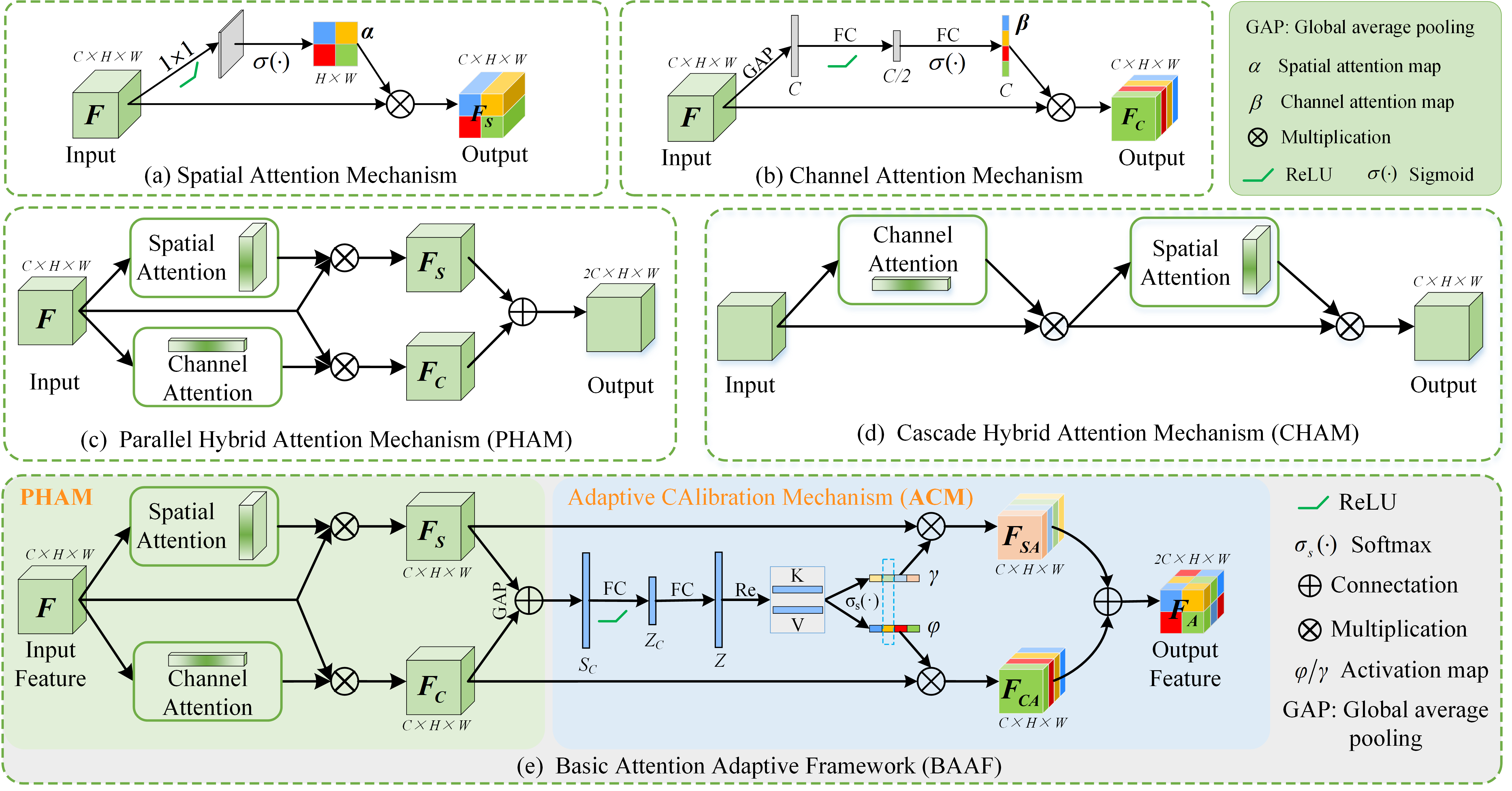}
	\begin{sloppypar}
		\textbf{Fig. 4.} The comparison of the proposed basic attentional adaptive framework and the existing attention mechanisms.
	\end{sloppypar}
\end{figure*}
%% 文件目录
\graphicspath{{Fig1/}} 
%% h：当前位置，t：顶端，b：下，p：浮动 ；ht，htbp组合
\begin{figure*}[ht]\footnotesize
	\centering
	\includegraphics[scale=0.84]{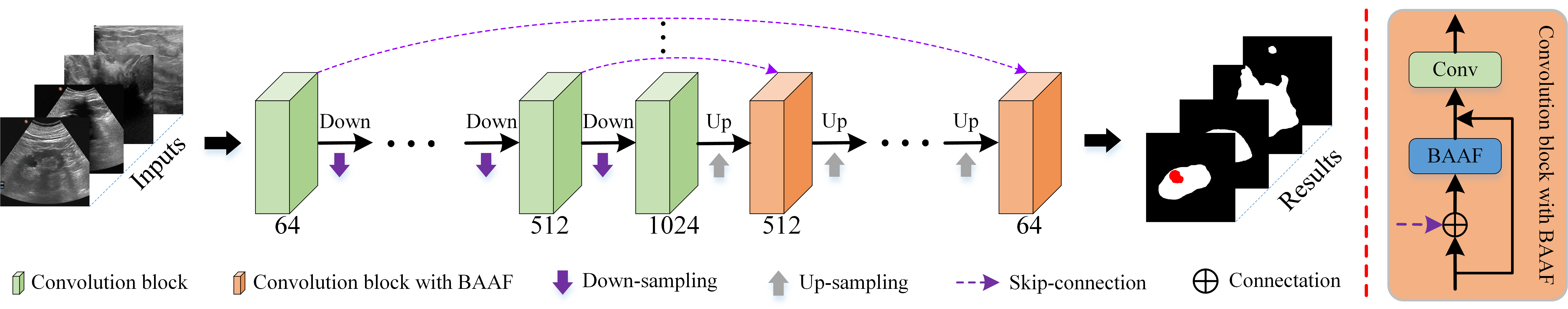}
	\begin{sloppypar}
		\textbf{Fig. 5.} The U-shaped network we constructed using the BAAF block.
	\end{sloppypar}
\end{figure*}

\section{Method}
\subsection{Basic Attention Adaptive Framework (BAAF)}
As shown in Fig. \textcolor{red}{4}, the BAAF has two main components: a parallel hybrid attention mechanism (PHAM) and an adaptive calibration mechanism (ACM). The PHAM is responsible for calibrating the input features from the channel and spatial dimensions, and the ACM is mainly focused on selecting more robust representations adaptively from the calibrated feature maps.

\subsubsection{Parallel Hybrid Attention Mechanism (PHAM)}
Inspired by the scSE attention \textcolor{blue}{\cite{Roy2018}}, we also adopted a parallel hybrid attention mechanism (PHAM) to calibrate the input features. Specially, the designed PHAM contains a channel attention module and a spatial attention module. The detailed descriptions of channel attention and spatial attention are shown in Fig. \textcolor{red}{4(a)} and Fig. \textcolor{red}{4(b)}, respectively. In the spatial attention module, the input feature map $F \in {{\mathbb{R}}^{c \times h \times w}}$ is first executed with a $1 \times 1$ convolution operation and a ReLU activation operation. Then, a sigmoid activity is performed to obtain the spatial dimension of the activation map $\alpha $.
\begin{equation}
	\alpha {\rm{ = }}\sigma ({\delta _r}({W_{1 \times 1}} \cdot F)),
\end{equation}
where ${W_{1 \times 1}}$ denotes the matrix of the $1 \times 1$ convolution operation. The ${\delta _r}( \cdot )$ and $\sigma ( \cdot )$ represent the ReLU activity and the sigmoid activity, respectively. The value in $\alpha $ emphasizes the importance of the corresponding spatial information in the feature map $F \in {{\mathbb{R}}^{c \times h \times w}}$ . Finally, $\alpha $ is expanded to the same dimension of the feature map $F \in {{\mathbb{R}}^{c \times h \times w}}$ and the calibration activity is executed on the feature map $F \in {{\mathbb{R}}^{c \times h \times w}}$. The calibrated features can be represented as: 
\begin{equation}
	{F_S}{\rm{ = }}Ex(\alpha ) \otimes F,
\end{equation}
where ${F_S} \in {{\mathbb{R}}^{c \times h \times w}}$ denotes the output of the spatial attention branch. The $Ex( \cdot )$ and $ \otimes $ represent the expansion activity and calibration activity, respectively.

In the channel attention module, the input feature map $F \in {{\mathbb{R}}^{c \times h \times w}}$ is first executed with a global average pooling (GAP) operation to compress the feature scale. Subsequently, the compressed feature map undergoes two fully connected operations, ReLU activation and sigmoid activation, obtain the channel attention map $\beta $.
\begin{equation}
	\beta {\rm{ = }}\sigma ({W_{f2}} \cdot {\delta _{\rm{r}}}({W_{f1}} \cdot GAP(F))),
\end{equation}
where ${W_{f1}} \in {{\mathbb{R}}^{{\textstyle{c \over r}} \times c}}$ and ${W_{f2}} \in {{\mathbb{R}}^{c \times {\textstyle{c \over r}}}}$ represent the matrix of two fully connected layers, respectively. Similarly, The value in $\beta $ emphasizes the importance of the corresponding channel information in the feature map $F \in {{\mathbb{R}}^{c \times h \times w}}$. Finally, $\beta $ is reshaped to the same scale of the feature map $F \in {{\mathbb{R}}^{c \times h \times w}}$ and the calibration activity is executed on the feature map $F \in {{\mathbb{R}}^{c \times h \times w}}$. The calibrated features can be represented as: 
\begin{equation}
	{F_C}{\rm{ = }}Re(\beta ) \otimes F,
\end{equation}
where ${F_C} \in {{\mathbb{R}}^{c \times h \times w}}$ denotes the output of the channel attention model. The $Re( \cdot )$ represents the reshape activity.

\subsubsection{Adaptive Calibration Mechanism (ACM)}
Hybrid attention mechanisms can give different perspectives to regions of interest (ROI) compared to individual attention mechanisms, but simply aggregating calibrated features from multiple attention mechanisms still has some limitations \textcolor{blue}{\cite{Roy2018, Woo2018}}. To capture more robust objective characterizations from different attention mechanisms, we proposed an adaptive calibration mechanism (ACM) for adaptive selection of feature information from different attentions. Specially, we first aggregate global information from different attention-calibrated features by simply using global average pooling (GAP) to generate channel-wise statistics as ${S_C} \in {{\mathbb{R}}^{c \times 1}}$.
\begin{equation}
	{S_C} = GAP({F_S}) \oplus GAP({F_C}),
\end{equation}
where $GAP( \cdot )$ represents the global average pooling operation, add $ \oplus $ denotes the add action. Then, the full convolution operation, the ReLU activation and the full convolution operation are performed on the feature vector ${S_C}$ to produce a new set of feature vectors:
\begin{equation}
	Z = {W_{fc2}} \cdot ({\delta _r}({W_{fc1}} \cdot {S_C})),
\end{equation}
where ${W_{fc2}} \in {{\mathbb{R}}^{2C \times 1}}$ is the matrix of the second full convolution operation, ${\delta _r}( \cdot )$ denotes the ReLU operation, and ${W_{fc1}} \in {{\mathbb{R}}^{d \times 1}}$ represents the matrix of the first full convolution operation. $d$ is squeezed dimension, which can be shown as:
\begin{equation}
	d = \max ({C \mathord{\left/
			{\vphantom {C r}} \right.
			\kern-\nulldelimiterspace} r},{\rm{ }}32),
\end{equation}
where $r$ is the reduction rate with default value of 8. Finally, $Z$ is executed the reshape operation and the softmax activation ${\sigma _s}( \cdot )$ to generate the channel-wise activation maps of ${F_C}$ and ${F_S}$, respectively. These channel-wise activation maps can be denoted as:
\begin{equation}
	\varphi  = {{{e^K}} \over {{e^K} + {e^V}}},
\end{equation}
\begin{equation}
	\gamma  = {{{e^V}} \over {{e^K} + {e^V}}},
\end{equation}
\begin{equation}
	{\varphi _i} + {\gamma _i} = 1,
\end{equation}
where $K$ and $V \in {{\mathbb{R}}^{C \times 1}}$ denote the feature vector after the reshape action. $\varphi $ and $\gamma  \in {{\mathbb{R}}^{C \times 1}}$ represent the channel activation maps of ${F_C}$ and ${F_S}$. Note that ${\varphi _i}$ is the $i - th$ element of $\varphi $. Subsequently, we merged the feature maps calibrated by $\varphi $ and $\gamma $.
\begin{equation}
	{F_A} = \varphi  \cdot {F_C} \oplus \gamma  \cdot {F_S},
\end{equation}
where ${F_A} \in {{\mathbb{R}}^{2C \times H \times W}}$ is the output feature map of the BAAF module, and $ \oplus $ denotes the concatenate action. 

\subsection{The overall U-shaped Framework}
Currently, U-net and its variants are widely and successfully used for medical image segmentation tasks \textcolor{blue}{\cite{Chen2023b}}. Inspired by this, we first utilize a U-net with the depth of 15 as the benchmark architecture. Then, the proposed BAAF module is added to the benchmark network to construct a new U-shaped framework, as shown in Fig. \textcolor{red}{5}. The U-shaped network contains seven down-sampling and seven up-sampling operations. During the encoding process, each convolutional module is composed of two $3 \times 3$ convolutional layers, two batch normalization layers, and two LeakyReLU layers. In the decoding stage, the BAAF components are introduced into each convolutional module. The filter size of the U-shaped network is 64, 128, 128, 256, 256, 512, 512, 1024, 512, 512, 256, 256, 128, 128, 64.

%\url{http://www.elsevier.com/wps/find/journaldescription.cws_home/505619/authorinstructions}).
\section{Materials and Experiments}
\subsection{Medical ultrasound datasets}
In this work, four medical ultrasound datasets are used to evaluate the performance of the proposed method. The first ultrasound dataset (Named: Dataset 1) was constructed by Al-Dhabyani et al., \textcolor{blue}{\cite{Al-Dhabyani2020}}. Dataset 1 includes 780 images captured by two ultrasound devices (LOGIQ E9 and LOGIQ E9 Agile) at Baheya Hospital. The average size of these images is $500 \times 500$ pixels. The second ultrasound dataset (Named: Dataset 2) was collected and obtained by Yap et al., \textcolor{blue}{\cite{Yap2020}}. Dataset 2 includes 163 images collected by the Siemens ACUSON Sequoia C512 system, the average image size is $760 \times 570$ pixels. Dataset 1 and Dataset 2 are described in more detail in references \textcolor{blue}{\cite{Al-Dhabyani2020}} and \textcolor{blue}{\cite{Yap2020}}, respectively. 

The third and fourth ultrasound datasets (Named: Dataset 3 and Dataset 4) were collected with the assistance of the General Hospital of the Chinese People's Liberation Army and the Civil Aviation General Hospital. Dataset 3 contains 300 kidney ultrasound images are collected by Esaote MyLab and Philips EPIQ7 ultrasound devices, which is used only for kidney segmentation. In Dataset 3, the cases acquired by single ultrasound device are 150. Dataset 4 includes 300 ultrasound images acquired by Esaote MyLab, Hitachi, and Philips EPIQ7 ultrasound machines, which is performed for kidney and cyst segmentation. In Dataset 4, the cases acquired by single ultrasound device are 100. The segmentation masks in Dataset 3 and Dataset 4 were obtained by two experienced doctors manually labeled, respectively. \textcolor{red}{Once the manuscript is accepted, we will publish two kidney ultrasound datasets with the paper. Currently, we have obtained permission from the hospital to release these data. As these data are used by more researchers, it will further accelerate the process of intelligent diagnosis in clinical ultrasound.} In this study, these medical ultrasound images were resize to $384 \times 384$ for training and testing of the network.

\begin{table*}[t]\footnotesize
	\textbf{Table 1}
	\begin{sloppypar}
		The quantitative evaluation results (mean ± std) on Dataset 1 and Dataset 2. The top 2 scores are marked with bold and green text, respectively. “\textcolor[rgb]{ 0,  .69,  .94}{†}” represents the medicine-specific method.
	\end{sloppypar}
	\setlength{\parskip}{0.6em}
	\centering
	\tabcolsep=1.7mm %左右间距
	\renewcommand\arraystretch{1.3} %上下间距
	\begin{tabular}{c|ccccc|ccccc}
		\hline
		\multicolumn{1}{c|}{} & \multicolumn{5}{c|}{\textbf{Dataset 1 (Breast ultrasound)}}    & \multicolumn{5}{c}{\textbf{Dataset 2 (Breast ultrasound)}} \\
		\hline
		Methods & Dice & Jaccard & Recall & Specificity & Precision & Dice & Jaccard & Recall & Specificity & Precision \\
		\hline
		U-net & 70.10±2.2 & 60.70±2.36 &76.30±2.48 & 96.18±0.55 & 71.88±2.41 & 68.20±4.23 & 58.44±4.26 & 75.32±2.85 & 98.44±0.40 & 70.27±6.11 \\
		Att U-net & 67.99±1.18 & 57.09±1.22 & 66.97±4.08 & 96.87±0.83 & 78.78±4.67 & 69.30±4.07 & 59.93±4.53 & 76.15±4.21 & 98.43±0.33 & 70.40±6.05 \\
		UNETR & 72.99±1.67 & 63.86±1.78 & \textbf{80.21±1.60} & 96.65±0.65 & 73.44±2.85 & 69.64±5.28 & 59.11±5.76 & 78.17±4.31 & 98.04±0.53 & 70.39±3.15 \\
		U-net++ & 71.58±2.09 & 61.38±1.73 & 71.44±2.77 & 97.04±0.54 & \textbf{79.68±3.07} & 69.77±5.30 & 61.19±5.86 & 79.64±3.84 & 98.44±0.41 & 68.32±5.73 \\
		SegNet & 75.64±1.80 & 67.31±1.87 & 79.85±1.03 & 96.99±0.53 & 76.09±2.00 & 72.16±1.52 & 62.83±2.20 & \textbf{80.15±3.90} & 98.59±0.30 & 71.72±1.70 \\
		STAN\textcolor[rgb]{ 0,  .69,  .94}{\textbf{†}} & 73.04±2.95 & 64.10±3.05 & 78.39±2.16 & 96.64±0.67 & 73.96±3.30 & 66.06±4.24 & 57.09±3.92 & 69.95±6.17 & 98.58±0.47 & 67.71±3.11 \\
		RDAU-net\textcolor[rgb]{ 0,  .69,  .94}{\textbf{†}} & 71.94±3.46 & 63.75±3.36 & 78.90±1.35 & 96.63±0.76 & 71.25±4.11 & 68.22±4.94 & 58.17±4.91 & 73.55±5.28 & 98.37±0.39 & 70.49±4.26 \\
		MADU-net\textcolor[rgb]{ 0,  .69,  .94}{\textbf{†}} & 71.35±2.67 & 61.62±2.69 & 76.87±2.58 & 96.40±0.62 & 73.77±2.90 & 72.32±3.14 & 63.09±3.04 & 79.24±1.72 & 98.61±0.36 & 73.70±5.08\\
		AE U-net\textcolor[rgb]{ 0,  .69,  .94}{\textbf{†}} & 73.47±3.03 & 64.57±2.91 & 79.00±2.11 & 96.80±0.54 & 74.44±3.74 & 72.23±2.14 & 62.37±2.16 & 78.97±2.29 & 98.67±0.28 & 72.27±1.91 \\
		SKU-net\textcolor[rgb]{ 0,  .69,  .94}{\textbf{†}} & \textbf{76.92±1.57} & \textbf{68.10±1.63} & 79.53±1.93 & \textbf{97.33±0.45} & 78.62±1.66 & \textbf{73.53±4.05} & \textbf{64.25±4.01} & 79.36±2.50 & \textbf{98.68±0.39} & \textbf{75.27±6.70} \\
		Ours  & \textcolor[rgb]{ 0,  .69,  .314}{\textbf{79.05±2.24}} & \textcolor[rgb]{ 0,  .69,  .314}{\textbf{70.67±2.34}} & \textcolor[rgb]{ 0,  .69,  .314}{\textbf{83.64±2.56}} & \textcolor[rgb]{ 0,  .69,  .314}{\textbf{97.47±0.59}} & \textcolor[rgb]{ 0,  .69,  .314}{\textbf{79.79±1.68}} & \textcolor[rgb]{ 0,  .69,  .314}{\textbf{80.85±1.48}} & \textcolor[rgb]{ 0,  .69,  .314}{\textbf{72.68±2.15}} & \textcolor[rgb]{ 0,  .69,  .314}{\textbf{84.49±2.36}} & \textcolor[rgb]{ 0,  .69,  .314}{\textbf{98.96±0.28}} & \textcolor[rgb]{ 0,  .69,  .314}{\textbf{80.77±3.17}} \\
		P-value & 3.83e-06 & 5.05e-07 & 3.13e-06 & 4.95e-2 & 3.95e-3 & 3.03e-05 & 5.24e-06 & 2.26e-02 & 4.78e-02 & 1.57e-03 \\
		\hline
	\end{tabular}
\end{table*}

\subsection{Experimental settings}
To ensure the fairness and reliability of the comparison experiments, K-fold cross-validation was implemented on the four ultrasound datasets. Specifically, four-fold cross-validation was performed on Dataset 1 and Dataset 2. Three-fold cross-validation was applied to Dataset 3 and Dataset 4, respectively. The previous study demonstrated that most of the methods have better segmentation performance on kidney ultrasound images than breast ultrasound images \textcolor{blue}{\cite{Chen2023}}. Compared with the four-fold cross-validation experiments on the breast ultrasound dataset, the three-fold cross-validation experiments on the kidney ultrasound dataset not only can assess the robustness of different segmentation methods in terms of cross-validation experimental ways, but also can further reduce the network training time consumption. During training, the binary cross entropy (BCE) is used as loss function of the network. The Adam (initial learning rate is 1e-3) was chosen as the optimizer of the network. The development environment of our network is Ubuntu 20.04, python 3.6, TensorFlow 2.6.0, NVIDIA RTX 3090 GPUs. In the cross-validation experiment, we randomly select 20$\%$ from the training data of each fold to be considered as validation data to determine the conditions for network termination. This differs from typical K-fold cross-validation experiments, but does not affect the fairness of the comparison experiments. The selection of randomised validation data is more representative of the generalisability of the segmentation method. Finally, the epoch size and batch size are set to 50 and 12, respectively.

%% 文件目录
\graphicspath{{Fig1/}} 
%% h：当前位置，t：顶端，b：下，p：浮动 ；ht，htbp组合
\begin{figure*}[!ht]\footnotesize
	\centering
	\includegraphics[scale=.3]{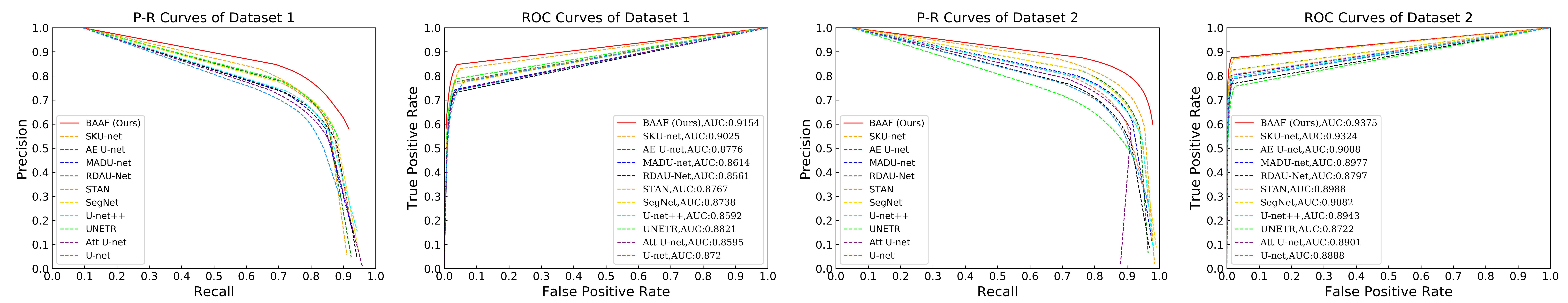}
	\begin{sloppypar}
		\textbf{Fig. 6.} P-R curves and ROC curves of different methods on Dataset 1 and Dataset 2.
	\end{sloppypar}
\end{figure*}

\subsection{Evaluation metrics}
To adequately demonstrate the capability of the proposed method on medical ultrasound segmentation tasks, eight quantitative evaluation metrics are applied to evaluate the difference between the predicted results and the ground-true masks. They are composed of area-based evaluation metrics Jaccard, Precision, Recall, Specificity, Dice and boundary-based evaluation metrics hausdorff distance (HD), average symmetric surface distance (ASSD) and average boundary distance (ABD) \textcolor{blue}{\cite{Wang2020b}}. The complexity of breast ultrasound can cause the segmentation network to miss-recognize the whole image as background. Therefore, boundary-based evaluation metrics cannot be used to evaluate the segmentation performance of the segmentation network on breast ultrasound.

%% 文件目录
\graphicspath{{Fig1/}} 
%% h：当前位置，t：顶端，b：下，p：浮动 ；ht，htbp组合
\begin{figure*}[!ht]\footnotesize
	% \centering
	\includegraphics[scale=.375]{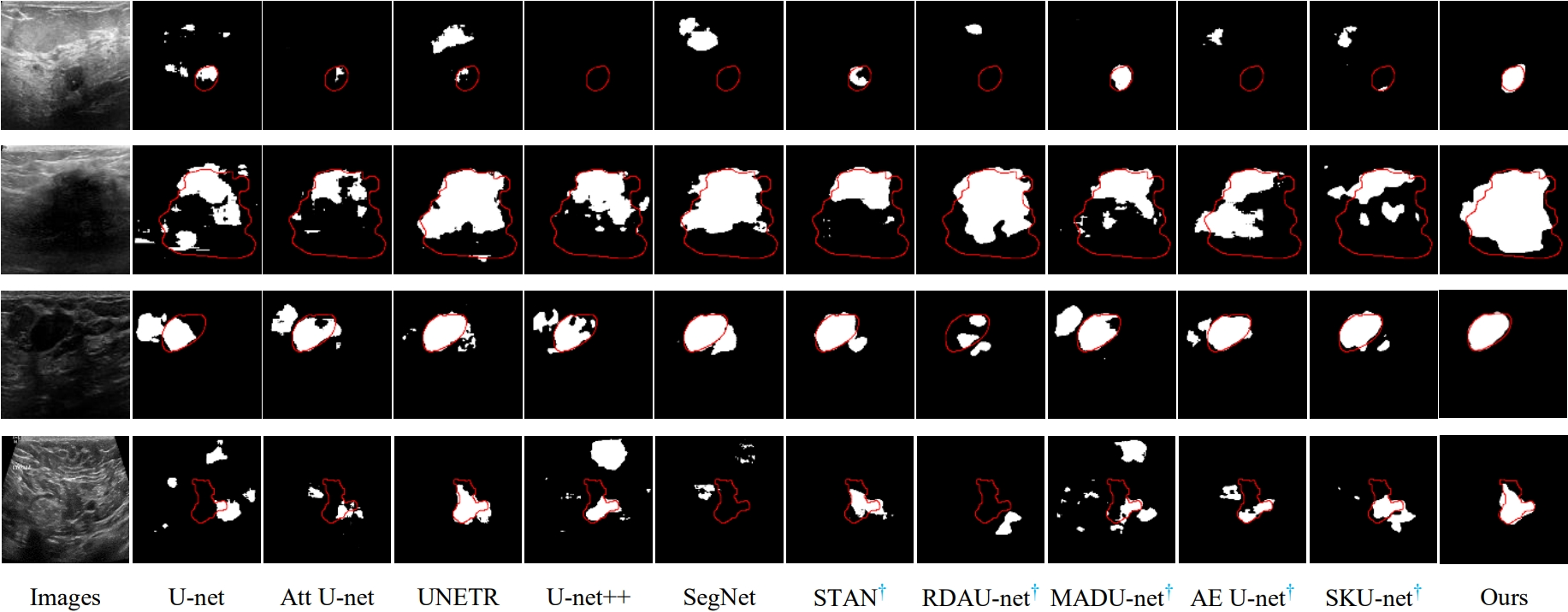}
	\begin{sloppypar}
		\textbf{Fig. 7.} The visualization of prediction mask of different methods on Dataset 1 and Dataset 2. “\textcolor[rgb]{ 0,  .69,  .94}{†}” represents the medicine-specific method. The red curve indicates the ground-true contour.
	\end{sloppypar}
\end{figure*}

\begin{table*}[!t]\footnotesize
	\textbf{Table 2}
	\begin{sloppypar}
		The quantitative evaluation results (mean ± std) on Dataset 3 and Dataset 4. The top 2 scores are marked with bold and green text, respectively. “\textcolor[rgb]{ 0,  .69,  .94}{†}” represents the medicine-specific method.
	\end{sloppypar}
	\setlength{\parskip}{0.5em}
	\centering
	\tabcolsep=0.05mm %左右间距
	\renewcommand\arraystretch{1.3} %上下间距
	\begin{tabular}{c|cccccccccccc}
		\hline
		Methods & U-net & Att U-net  & UNETR & U-net++ & SegNet & AE U-net & SKU-net & SDFNet\textcolor[rgb]{ 0,  .69,  .94}{\textbf{†}} & MSDSNet\textcolor[rgb]{ 0,  .69,  .94}{\textbf{†}} & MBANet\textcolor[rgb]{ 0,  .69,  .94}{\textbf{†}} & Ours  & P-value \\
		\hline
		Jaccard & 85.00±1.39 & 85.99±1.50 & 81.42±1.67 & 85.59±1.53 & 85.78±0.98 & 83.90±4.03 & 87.16±0.42 & 85.81±1.94 & 87.28±0.80 & \textbf{88.03±1.28} & \textcolor[rgb]{ 0,  .69,  .314}{\textbf{89.25±0.61}} & 2.8e-04 \\
		Precision & 93.03±1.86 & 91.56±1.99 & 88.22±2.31 & 90.33±1.69 & 92.66±1.47 & 91.98±1.22 & 93.23±1.36 & 90.74±3.12 & \textbf{93.90±1.79} & 93.30±1.65 & \textcolor[rgb]{ 0,  .69,  .314}{\textbf{94.38±1.51}} & 4.8e-02 \\
		Recall & 91.11±0.79 & 93.70±0.32 & 91.44±1.46 & 93.22±0.22 & 92.31±0.89 & 90.91±5.88 & 93.36±0.93 & \textbf{94.43±1.00} & 92.84±1.36 & 94.03±0.61 & \textcolor[rgb]{ 0,  .69,  .314}{\textbf{94.52±1.10}} & 4.9 e-02 \\
		Specificity & 98.77±0.60 & 98.50±0.46 & 97.91±0.65 & 98.28±0.51 & 98.67±0.63 & 98.54±0.44 & 98.72±0.37 & 98.35±0.63 & \textbf{98.82±0.57} & 98.70±0.57 & \textcolor[rgb]{ 0,  .69,  .314}{\textbf{99.00±0.28}} & 4.8e-02 \\
		Dice  & 91.57±0.86 & 92.10±0.92 & 89.20±1.15 & 91.78±1.01 & 91.90±0.60 & 90.79±2.61 & 92.91±0.25 & 91.88±1.39 & 92.91±0.55 & \textbf{93.35±0.87} & \textcolor[rgb]{ 0,  .69,  .314}{\textbf{94.20±0.36}} & 4.0e-04 \\	
		HD    & 57.38±17.83 & 52.14±8.22 & 82.73±5.34 & 57.42±4.92 & 42.43±13.41 & 53.32±33.08 & 31.57±9.00 & 31.26±8.50 & \textbf{23.79±6.57} & 24.27±11.03 & \textcolor[rgb]{ 0,  .69,  .314}{\textbf{11.91±1.72}} & 1.4e-06 \\
		ASSD  & 2.09±0.99 & 1.03±0.04 & 1.95±0.77 & 0.79±0.21 & \textbf{0.71±0.04} & 2.17±2.42 & 0.95±0.27 & 0.87±0.17 & 1.07±0.69 & 0.72±0.09 & \textcolor[rgb]{ 0,  .69,  .314}{\textbf{0.42±0.07}} & 1.8e-03 \\
		ABD   & 9.62±3.43 & 6.39±0.97 & 10.36±1.63 & 6.28±0.81 & 5.03±0.85 & 8.36±5.13 & 5.33±0.55 & 5.84±0.59 & 5.16±1.21 & \textbf{4.24±0.68} & \textcolor[rgb]{ 0,  .69,  .314}{\textbf{3.48±0.29}} & 6.3e-05 \\
		\hline
		Jaccard & 71.34±0.88 & 72.15±1.71 & 51.60±4.63 & 74.93±2.96 & 79.94±1.52 & 79.43±1.38 & 80.51±1.33 & 73.13±1.57 & \textbf{82.19±0.63} & 81.05±1.69 & \textcolor[rgb]{ 0,  .69,  .314}{\textbf{84.97±1.25}} & 5.0e-07 \\
		Precision & 82.58±1.00 & 80.40±1.76 & 66.90±7.03 & 81.20±2.86 & 87.19±1.99 & 86.37±1.77 & 88.41±0.34 & 80.23±1.47 & \textbf{89.66±0.72} & 87.27±2.37 & \textcolor[rgb]{ 0,  .69,  .314}{\textbf{91.85±0.47}} & 5.5e-05 \\
		Recall & 84.58±0.96 & 88.58±0.81 & 72.43±7.01 & 91.45±0.67 & 90.89±0.34 & 91.08±0.73 & 90.56±1.49 & 89.74±0.80 & 91.32±1.01 & \textbf{91.56±0.42} & \textcolor[rgb]{ 0,  .69,  .314}{\textbf{92.43±1.01}} & 1.4e-02 \\
		Specificity & 97.75±0.05 & 97.48±0.20 & 95.74±0.71 & 97.63±0.33 & 98.40±0.11 & 98.25±0.14 & 98.45±0.09 & 97.54±0.14 & \textbf{98.65±0.07} & 98.42±0.13 & \textcolor[rgb]{ 0,  .69,  .314}{\textbf{98.92±0.09}} & 1.0e-04 \\
		Dice  & 82.03±0.70 & 82.52±1.44 & 66.49±4.12 & 84.01±2.35 & 87.32±1.56 & 87.39±1.34 & 88.41±0.92 & 82.76±1.34 & \textbf{89.31±0.58} & 88.12±1.51 & \textcolor[rgb]{ 0,  .69,  .314}{\textbf{91.29±0.87}} & 2.0e-05 \\
		HD    & 104.93±13.90 & 118.88±5.76 & 166.11±21.46 & 101.25±10.87 & 70.42±13.33 & 61.28±9.24 & 63.08±11.26 & 99.77±1.74 & 51.30±5.58 & \textbf{48.85±4.97} & \textcolor[rgb]{ 0,  .69,  .314}{\textbf{25.59±4.88}} & 4.7e-10 \\
		ASSD  & 5.82±0.66 & 4.22±0.76 & 13.16±5.41 & 2.68±0.25 & 2.51±0.23 & 2.60±0.37 & 2.35±0.53 & 3.77±0.21 & 2.09±0.55 & \textbf{2.09±0.43} & \textcolor[rgb]{ 0,  .69,  .314}{\textbf{1.64±0.35}} & 3.2e-02 \\
		ABD   & 19.01±1.57 & 18.58±1.68 & 32.43±4.72 & 13.59±0.91 & 9.18±0.72 & 10.04±0.74 & 10.00±0.50 & 15.34±0.49 & 8.74±0.43 & \textbf{7.93±0.34} & \textcolor[rgb]{ 0,  .69,  .314}{\textbf{6.67±0.55}} & 7.2e-05 \\
		\hline	
	\end{tabular}
\end{table*}

%% 文件目录
\graphicspath{{Fig1/}} 
%% h：当前位置，t：顶端，b：下，p：浮动 ；ht，htbp组合
\begin{figure*}[!ht]\footnotesize
	\centering
	\includegraphics[scale=.3]{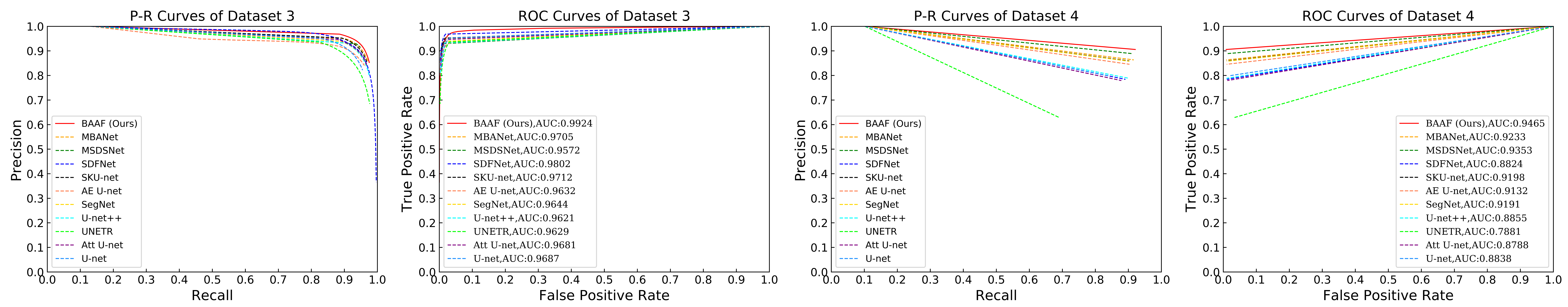}
	\begin{sloppypar}
		\textbf{Fig. 8.} P-R curves and ROC curves of different methods on Dataset 3 and Dataset 4.
	\end{sloppypar}
\end{figure*}

\section{Experimental results and discussion}
In this paper, U-net \textcolor{blue}{\cite{Ronneberger2015}}, SegNet \textcolor{blue}{\cite{Badrinarayanan2017}}, Att U-net \textcolor{blue}{\cite{Oktay2018}}, U-net++ \textcolor{blue}{\cite{Zhou2020}}, UNETR \textcolor{blue}{\cite{Hatamizadeh2022}} five classical medical segmentation networks are used for comparison experiments. For the comparison of breast ultrasound, we also selected five specific segmentation methods, AE U-Net \textcolor{blue}{\cite{Yan2022}}, STAN \textcolor{blue}{\cite{Shareef2020}}, RDAU-Net \textcolor{blue}{\cite{Zhuang2019b}}, MADU-net \textcolor{blue}{\cite{Abraham2019}}, SKU-net \textcolor{blue}{\cite{Byra2020}}. Similarly, we used three specific methods SDFNet \textcolor{blue}{\cite{Chen2021}}, MBANet \textcolor{blue}{\cite{105140}}, and MSDSNet \textcolor{blue}{\cite{chen2022novel}} for kidney ultrasound. According to the open-source codes, we retrain these comparison networks on the same datasets as our method. To ensure absolute objectivity in comparison experiments, the prediction results for all methods are not subjected to any post-processing.

\subsection{Comparison results on breast ultrasound}
Table \textcolor{red}{1} illustrates the quantitative evaluation results of breast ultrasound. Fig. \textcolor{red}{6} shows the P-R curve and ROC curve of these comparison methods on Dataset 1 and Dataset 2. Fig. \textcolor{red}{7} presents the visualized predicted masks of different methods on breast ultrasound. It is worth noting that there are some cases of segmentation failures on breast ultrasound as shown in Fig. \textcolor{red}{7}, and hence boundary-based evaluation metrics cannot be used. In Dataset 1, our method achieves the best results on five quantitative indicators, and their values are 79.05$\%$, 70.67$\%$, 83.64$\%$, 97.47$\%$ and 79.79$\%$. In Dataset 2, our method also achieved the best results on five quantitative indicators, which are 80.85$\%$, 72.68$\%$, 84.49$\%$, 98.96$\%$ and 80.77$\%$. To further highlight the improvement of our method on breast ultrasound for these quantitative metrics, we conducted the statistical analysis of t-test. As shown in Table \textcolor{red}{1}, the p-values ($p<0.05$) based on the t-test indicate that our method has a significant improvement on these quantitative metrics. As shown in Fig. \textcolor{red}{6}, our method achieved the best results on Dataset 1 and Dataset 2 in terms of AUC metrics. The area below the P-R curve and ROC curve can indicate the confidence level of a method, the larger the area means the confidence level is also high. Based on the P-R curve and ROC curve, we can conclude that the segmentation results of our proposed BAAF module on two ultrasound datasets have a higher confidence level. It can be seen from Fig. \textcolor{red}{7}, similar intensity distributions and confusing boundaries can cause serious missed and false detections, and even fail to localize the region of interest. Compared with the masks predicted by other methods, our method can reduce the occurrence of the missed and false detection. In summary, the method proposed in this paper achieves satisfactory results in the automatic segmentation of breast ultrasound.

%% 文件目录
\graphicspath{{Fig1/}} 
%% h：当前位置，t：顶端，b：下，p：浮动 ；ht，htbp组合
\begin{figure*}[!ht]\footnotesize
	% \centering
	\includegraphics[scale=.375]{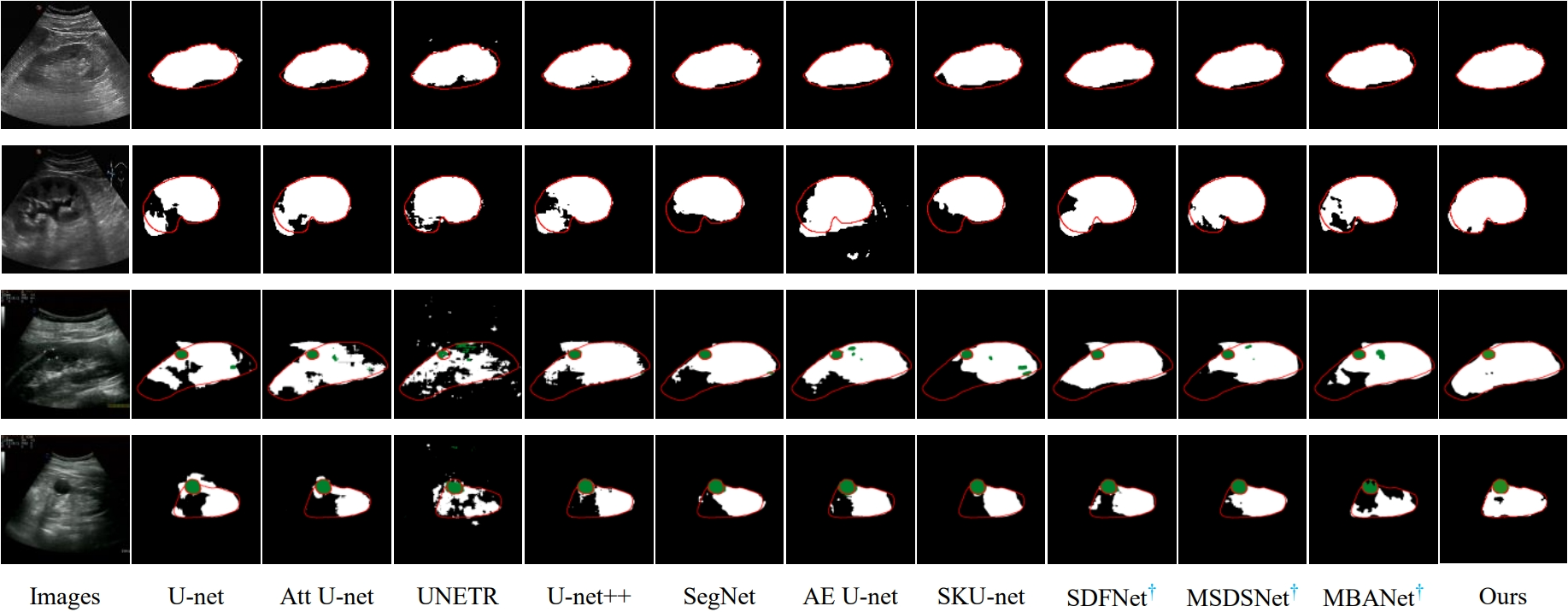}
	\begin{sloppypar}
		\textbf{Fig. 9.} The visualization of prediction mask of different methods on Dataset 3 and Dataset 4. “\textcolor[rgb]{ 0,  .69,  .94}{†}” represents the medicine-specific method. The red curve indicates the ground-true contour.
	\end{sloppypar}
\end{figure*}

\subsection{Comparison results on kidney ultrasound}
Table \textcolor{red}{2} illustrates the quantitative evaluation results of kidney ultrasound. As shown in Table \textcolor{red}{2}, our method achieves the best results on eight quantitative indicators. In Dataset 3, the values of the eight quantitative indicators are 89.25$\%$, 94.38$\%$, 94.52$\%$, 99.00$\%$, 94.20$\%$, 11.91, 0.43 and 3.48. Compared to the second result, five indicators (Jaccard, Precision, Recall, Specificity and Dice) of Dataset 3 are increased on average by 0.6$\%$ and three indicators (HD, ASSD, ABD) are reduced on average by 63.94$\%$. In Dataset 4, the values of the eight quantitative indicators are 84.97$\%$, 91.85$\%$, 92.43$\%$, 98.92$\%$, 91.29$\%$, 25.59, 1.64 and 6.67. Compared to the second result, five indicators (Jaccard, Precision, Recall, Specificity and Dice) of Dataset 3 are increased on average by 1.79$\%$ and three indicators (HD, ASSD, ABD) are reduced on average by 45.68$\%$. As shown in Table \textcolor{red}{2}, the p-values ($p<0.05$) based on the t-test indicate that our method has a significant improvement on these quantitative metrics. Fig. \textcolor{red}{8} presents the P-R curve and ROC curve of these comparison methods on Dataset 3 and Dataset 4. In terms of AUC metrics, our method achieved the best results on two datasets. According to the P-R curve and ROC curve, we can conclude that the segmentation results of our proposed BAAF module on two ultrasound datasets have a higher confidence level. Fig. \textcolor{red}{9} presents the visualized predicted masks of different methods on kidney ultrasound. It can be seen from Fig. \textcolor{red}{9}, similar intensity distributions and confusing boundaries can cause serious missed and false detections. Compared with the masks predicted by other methods, our method can reduce the occurrence of missed and false detections, and its prediction results are closest to the ground-true mask. In general, our method achieves satisfactory performance in kidney ultrasound segmentation tasks.

\begin{table*}[h]\footnotesize
	\textbf{Table 3}
	\begin{sloppypar}
		The quantitative evaluation results (mean ± std) of different attention models on Dataset 1 and Dataset 2.
	\end{sloppypar}
	\setlength{\parskip}{0.9em}
	\centering
	\tabcolsep=2mm %左右间距
	\renewcommand\arraystretch{1.3} %上下间距
	\begin{tabular}{c|ccccc|ccccc}
		\hline
		\multicolumn{1}{c|}{} & \multicolumn{5}{c|}{\textbf{Dataset 1 (Breast ultrasound)}}    & \multicolumn{5}{c}{\textbf{Dataset 2 (Breast ultrasound)}} \\
		\hline
		Methods & Jaccard & Precision & Recall & Specificity & Dice & Jaccard & Precision & Recall & Specificity & Dice\\
		\hline
		AGM   & 68.57±1.00 & 78.82±1.79 & 81.14±1.54 & 97.36±0.54 & 77.07±0.86 & 65.34±7.34 & 73.82±6.31 & 80.35±7.23 & 98.61±0.28 & 73.41±7.07 \\
		SAM   & 68.02±0.87 & 77.66±1.89 & 80.03±2.05 & 97.30±0.48 & 76.27±0.87 & 67.70±2.24 & 77.19±1.16 & 80.63±2.32 & 98.76±0.27 & 76.03±2.15 \\
		ECA   & 69.24±2.51 & 78.15±2.18 & 81.71±2.64 & 97.25±0.51 & 77.41±2.52 & 68.65±3.25 & 77.96±3.13 & 80.78±4.18 & 98.92±0.29 & 76.79±2.73 \\
		CBAM  & 69.71±1.98 & 79.14±1.50 & 82.02±2.93 & 97.31±0.63 & 78.23±1.91 & 71.04±2.44 & 80.68±1.61 & 83.97±4.16 & 98.92±0.23 & 79.35±0.86 \\
		scSE  & 69.95±2.25 & 79.26±1.74 & 81.70±3.82 & 97.39±0.51 & 78.16±2.38 & 70.24±1.57 & 79.97±4.14 & 82.81±3.66 & 98.86±0.23 & 78.79±2.28 \\
		SK    & 69.18±2.17 & 78.44±2.56 & 81.40±1.73 & 97.24±0.44 & 77.63±2.29 & 68.19±1.72 & 80.30±1.40 & 79.92±2.38 & 98.87±0.19 & 77.45±1.46 \\
		HAAM  & 69.00±2.33 & 78.43±3.37 & 80.93±2.09 & 97.14±0.60 & 77.42±2.19 & 68.40±4.65 & 78.15±3.01 & 81.41±3.78 & 98.81±0.22 & 77.22±4.05 \\
		Ours  & \textbf{70.67±2.34} & \textbf{79.79±1.68} & \textbf{83.64±2.56} & \textbf{97.47±0.59} & \textbf{79.05±2.24} & \textbf{72.68±2.15} & \textbf{80.77±3.1}7 & \textbf{84.49±2.36} & \textbf{98.96±0.28} & \textbf{80.85±1.48} \\
		\hline
	\end{tabular}
\end{table*}

\begin{table*}[!h]\footnotesize
	\textbf{Table 4}
	\begin{sloppypar}
		The quantitative evaluation results (mean ± std) of different attention models on Dataset 3 and Dataset 4.
	\end{sloppypar}
	\setlength{\parskip}{2em}
	\centering
	\tabcolsep=2.5mm %左右间距
	\renewcommand\arraystretch{1.3} %上下间距
	\begin{tabular}{p{14.5mm}|c|cccccccc}
		\hline
		{} & Methods & AGM   & SAM   & ECA   & CBAM  & scSE  & SK    & HAAM  & Ours \\
		\hline
		\multirow{8}{14.5mm}{\textbf{Dataset 3 (Kidney ultrasound)}} & Jaccard & 86.35±1.59 & 87.49±0.74 & 88.33±1.07 & 88.09±0.42 & 88.75±0.66 & 88.04±0.48 & 88.02±1.08 & \textbf{89.25±0.61} \\
		 & Precision & 92.47±2.36 & 92.81±1.61 & 93.64±1.50 & 93.29±1.47 & 94.04±1.05 & 93.44±1.90 & 93.67±1.89 & \textbf{94.38±1.51} \\
		 & Recall & 93.26±0.72 & 94.22±1.04 & 94.28±0.76 & 94.37±1.07 & 94.30±0.71 & 94.20±1.40 & 93.92±1.21 & \textbf{94.52±1.10} \\
		 & Specificity & 98.63±0.48 & 98.65±0.52 & 98.87±0.38 & 98.76±0.47 & 98.97±0.25 & 98.78±0.49 & 98.79±0.49 & \textbf{99.00±0.28} \\
		 & Dice  & 92.37±1.02 & 93.04±0.52 & 93.61±0.70 & 93.46±0.24 & 93.90±0.39 & 93.47±0.31 & 93.42±0.69 & \textbf{94.20±0.36} \\
		 & HD    & 13.02±2.24 & 22.55±6.31 & 18.81±5.57 & 16.60±1.03 & 16.08±5.14 & 19.35±3.09 & 33.78±17.82 & \textbf{11.91±1.72} \\
		 & ASSD  & 0.79±0.16 & 0.59±0.06 & 0.56±0.09 & 0.59±0.22 & 0.53±0.06 & 0.54±0.21 & 0.56±0.08 & \textbf{0.42±0.07} \\
		 & ABD   & 4.74±0.71 & 4.47±0.45 & 3.85±0.51 & 4.20±0.34 & 3.80±0.48 & 4.11±0.19 & 4.29±0.64 & \textbf{3.48±0.29} \\
		\hline
		\multirow{8}{14.5mm}{\textbf{Dataset 4 (Kidney-cyst ultrasound)}} & Jaccard & 77.01±1.83 & 82.09±2.20 & 83.48±1.38 & 83.33±1.61 & 83.66±1.26 & 83.82±1.02 & 83.44±1.23 & \textbf{84.97±1.25} \\
		 & Precision & 85.43±1.96 & 88.83±2.59 & 90.91±0.99 & 90.83±1.51 & 90.75±0.68 & 90.35±0.58 & 91.01±0.64 & \textbf{91.85±0.47} \\
		 & Recall & 89.04±2.26 & 91.92±0.97 & 91.33±1.24 & 91.54±0.44 & 91.98±1.17 & 92.40±1.12 & 91.51±0.83 & \textbf{92.43±}1.01 \\
		 & Specificity & 98.17±0.27 & 98.58±0.22 & 98.81±0.13 & 98.80±0.14 & 98.74±0.10 & 98.68±0.09 & 98.72±0.10 & \textbf{98.92±0.09} \\
		 & Dice  & 85.63±1.11 & 89.12±1.77 & 90.25±1.07 & 90.23±1.22 & 90.45±0.95 & 90.60±0.72 & 90.38±0.89 & \textbf{91.29±0.87} \\
	 	 & HD    & 38.50±8.98 & 50.66±2.81 & 38.72±5.20 & 44.22±10.08 & 29.57±6.82 & 36.70±4.11 & 57.41±2.13 & \textbf{25.59±4.88} \\
		 & ASSD  & 3.19±0.78 & 2.23±0.39 & 1.89±0.61 & 1.88±0.18 & 1.69±0.42 & \textbf{1.57±0.25} & 1.98±0.22 & 1.64±0.35 \\
		 & ABD   & 10.84±1.70 & 8.56±0.55 & 7.51±0.68 & 7.80±0.10 & 7.26±0.65 & 7.35±0.27 & 8.07±0.54 & \textbf{6.67±0.55} \\
		\hline	
	\end{tabular}
\end{table*}

%% 文件目录
\graphicspath{{Fig1/}} 
%% h：当前位置，t：顶端，b：下，p：浮动 ；ht，htbp组合
\begin{figure*}[!h]\footnotesize
	\centering
	\includegraphics[scale=.3]{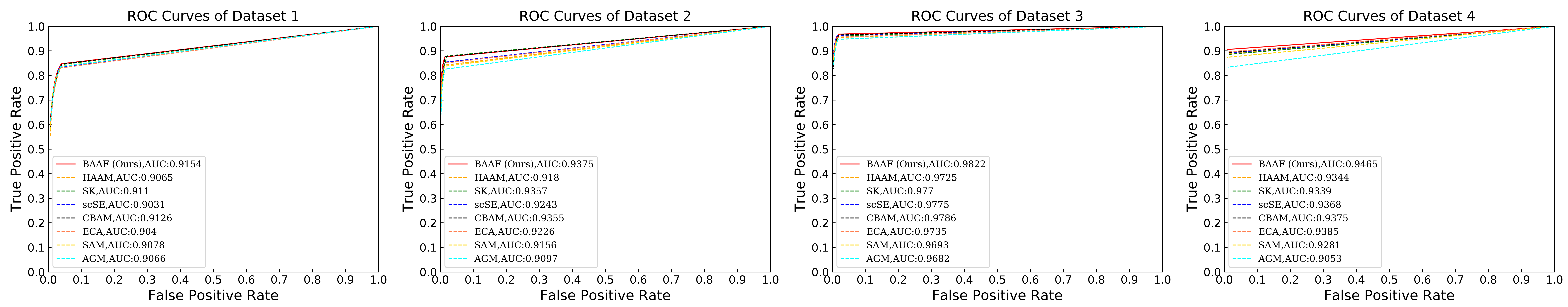}
	\begin{sloppypar}
		\textbf{Fig. 10.} The ROC curve of different attention modules on four medical ultrasound datasets.
	\end{sloppypar}
\end{figure*}

\begin{table*}[!h]\footnotesize
	%	\centering
	\textbf{Table 5}
	\begin{sloppypar}
		The segmentation results (mean ± std) of different framework components on Dataset 1 and Dataset 2. 
	\end{sloppypar}
	\setlength{\parskip}{2em}
	\centering
	\tabcolsep=2.6mm %左右间距
	\renewcommand\arraystretch{1.3} 
	\setlength{\tabcolsep}{0.8mm}{	
		\begin{tabular}{c|ccccc|ccccc}
			\hline
			{} & \multicolumn{5}{c|}{\textbf{Dataset 1 (Breast ultrasound)}}    & \multicolumn{5}{c}{\textbf{Dataset 2 (Breast ultrasound)}} \\
			\hline
			Methods & Jaccard & Precision & Recall & Specificity & Dice  & Jaccard & Precision & Recall & Specificity & Dice \\
			\hline
			U-net & 60.70±2.36 & 71.88±2.41 & 76.30±2.48 & 96.18±0.55 & 70.10±2.20 & 58.44±4.26 & 70.27±6.11 & 75.32±2.85 & 98.44±0.40 & 68.20±4.23 \\
			Deeper U-net & 68.91±1.88 & 78.70±2.79 & 81.50±2.31 & 97.36±0.51 & 77.31±1.95 & 67.86±1.87 & 76.59±2.76 & 80.95±2.31 & 98.65±0.55 & 76.48±1.77 \\
			Deep U-net + PHAM & 69.95±2.25 & 79.26±1.74 & 81.70±3.82 & 97.39±0.51 & 78.16±2.38 & 70.24±1.57 & 79.97±4.14 & 82.81±3.66 & 98.86±0.23 & 78.79±2.28 \\
			Deep U-net + BAAF (Ours) & \textbf{70.67±2.34} & \textbf{79.79±1.68} & \textbf{83.64±2.56} & \textbf{97.47±0.59} & \textbf{79.05±2.24} & \textbf{72.68±2.15} & \textbf{80.77±3.17} & \textbf{84.49±2.36} & \textbf{98.96±0.28} & \textbf{80.85±1.48} \\
			\hline
	\end{tabular}}
\end{table*}

\begin{table*}[!h]\footnotesize
	\textbf{Table 6}
	\begin{sloppypar}
		The segmentation results (mean ± std) of different framework components on Dataset 3 and Dataset 4. 
	\end{sloppypar}
	\setlength{\parskip}{2em}
	\centering
	\tabcolsep=2.6mm %左右间距
	\renewcommand\arraystretch{1.3} %上下间距
	\setlength{\tabcolsep}{1.6mm}{	
		\begin{tabular}{p{14.5mm}|c|cccccccc}
			\hline
			{} & Components & Jaccard & Precision & Recall & Specificity & Dice  & HD    & ASSD  & ABD \\
			\hline
			\multirow{4}{14.5mm}{\textbf{Dataset 3 (Kidney ultrasound)}} & U-net & 83.19±2.69 & 90.81±2.64 & 91.20±0.66 & 98.33±0.56 & 90.42±1.79 & 57.38±17.83 & 2.09±0.99 & 9.62±3.43 \\
			& Deep U-net & 88.22±0.77 & 93.81±1.69 & 93.97±1.02 & 98.83±0.45 & 93.55±0.50 & 18.84±3.07 & 0.58±0.03 & 4.12±0.43 \\
			& Deep U-net + PHAM & 88.75±0.66 & 94.04±1.05 & 94.30±0.71 & 98.97±0.25 & 93.90±0.39 & 16.08±5.14 & 0.53±0.06 & 3.80±0.48 \\
			& Deep U-net + BAAF (Ours) & \textbf{89.25±0.61} & \textbf{94.38±1.51} & \textbf{94.52±1.10} & \textbf{99.00±0.28} & \textbf{94.20±0.36} & \textbf{11.91±1.72} & \textbf{0.42±0.07} & \textbf{3.48±0.29} \\
			\hline
			\multirow{4}{14.5mm}{\textbf{Dataset 4 (Kidney-cyst ultrasound)}} & U-net & 71.34±0.88 & 82.58±1.00 & 84.58±0.96 & 97.75±0.05 & 82.03±0.70 & 104.93±13.90 & 5.82±0.66 & 19.01±1.57 \\
			& Deep U-net & 83.02±1.01 & 90.65±0.56 & 91.13±1.26 & 98.74±0.01 & 89.92±0.72 & 40.89±1.09 & 2.34±0.44 & 8.36±0.47 \\
			& Deep U-net + PHAM & 83.66±1.26 & 90.75±0.68 & 91.98±1.17 & 98.74±0.10 & 90.45±0.95 & 29.57±6.82 & 1.69±0.42 & 7.26±0.65 \\
			& Deep U-net + BAAF (Ours) & \textbf{84.97±1.25} & \textbf{91.85±0.47} & \textbf{92.43±1.01} & \textbf{98.92±0.09} & \textbf{91.29±0.87} & \textbf{25.59±4.88} & \textbf{1.64±0.35} & \textbf{6.67±0.55} \\
			\hline
	\end{tabular}}
\end{table*}

\subsection{Comparison with the attention module}
We presented comparative experiments with seven classical attention modules on four ultrasound datasets to further discuss the capabilities of the BAAF module. These attention networks include scSE \textcolor{blue}{\cite{Roy2018}}, CBAM \textcolor{blue}{\cite{Woo2018}}, SK \textcolor{blue}{\cite{Byra2020}}, HAAM \textcolor{blue}{\cite{Chen2023}}, AGM \textcolor{blue}{\cite{Zhang2019}}, SAM \textcolor{blue}{\cite{Zhong2020}} and ECA \textcolor{blue}{\cite{9156697}}. Among them, scSE and CBAM are two classic combinations of channel attention and spatial attention. HAAM is the variant based on SK, which aim to alleviate the impact of variable morphologies on network performance. Although hybrid attention (such as CBAM and scSE) achieved more competitive results on Dataset 1 and Dataset 2, their advantages on Dataset 3 and Dataset 4 were not significant. From the quantitative results presented in Table \textcolor{red}{3} and Table \textcolor{red}{4}, it can be concluded that AGM, SAM and ECA could not achieve satisfactory performance on medical ultrasound segmentation tasks. It is worth noting that although CBAM, scSE and SK obtained the best results on individual indicators, their overall performance needs to be further improved. In summary, the robustness of the proposed method in this paper is further illustrated through comparison with these state-of-the-art attention methods. Fig. \textcolor{red}{10} presents the ROC curve of different attention modules on four ultrasound datasets. Our method obtains the best AUC on four ultrasound datasets, which indicates that the segmentation results of the BAAF module have a higher confidence level than the other attention modules.

\subsection{Architecture ablation}
In this section, we conduct extensive experiments on four ultrasound datasets to evaluate the effectiveness of our proposed networks. In experiment, the original U-net with the depth of 9 is as the baseline. Subsequently, we demonstrate the performance of segmentation for the deeper U-net with the depth of 15. Why the depth of U-net is set to 15 has been described in detail in our existing work \textcolor{blue}{\cite{Chen2023b}}. Then, the parallel hybrid attention mechanism (PHAM) is added to the deep U-net. Finally, we present the segmentation performance after applying the attention adaptive framework (BAAF) to the deeper U-net. Table \textcolor{red}{5} and Table \textcolor{red}{6} demonstrate the quantitative results of the different network components on the four ultrasound datasets. As shown in Table \textcolor{red}{5} and Table \textcolor{red}{6}, increasing the depth of the network, introducing the PHAM module and using the BAAF module can further improve the segmentation accuracy of medical ultrasound images. Through comparing "Deep U-net" and "Deep U-net + PHAM", we can conclude that the introduction of the attention mechanism can enhance the attention to the region of interest and improve the performance of the network in the ultrasound segmentation task. According to the results of "Deep U-net + PHAM" and "Deep U-net + BAAF (Ours)", we can find that the BAAF module can further help the network to capture more robust object characteristics from the feature maps calibrated by the PHAM module. The visualized results of the different network components on the four datasets are displayed in Fig. \textcolor{red}{11}. As shown in Fig. \textcolor{red}{11}, the introduction of different components can make the predicted results closer to the ground-truth mask. In addition, the miss-detection and false-detection in the prediction results are further mitigated.

%% 文件目录
\graphicspath{{Fig1/}} 
%% h：当前位置，t：顶端，b：下，p：浮动 ；ht，htbp组合
\begin{figure}[!h]\footnotesize
%	\centering
	\centerline{\includegraphics[scale=.74]{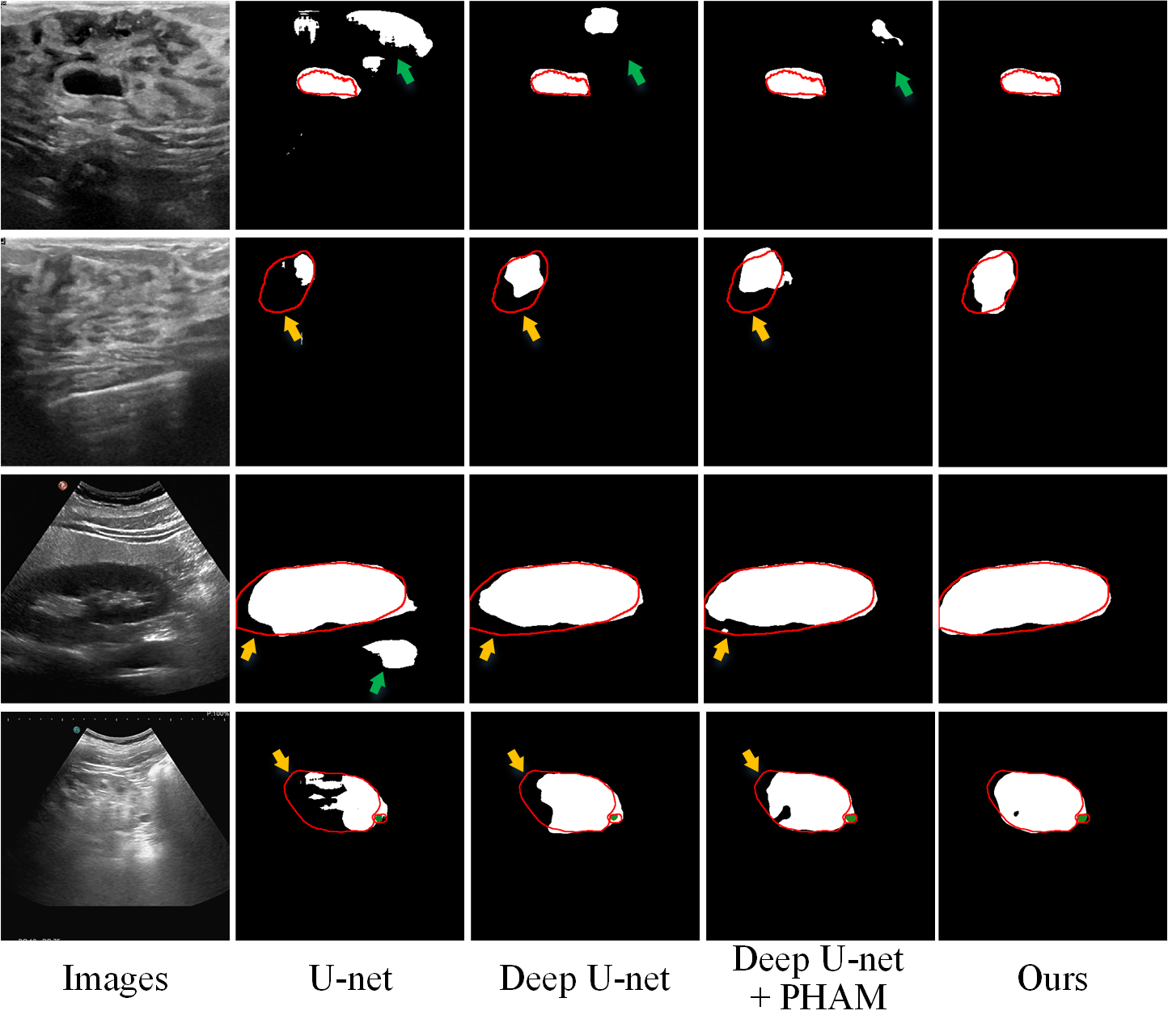}}
	\begin{sloppypar}
		\textbf{Fig. 11.} The prediction mask of different network components. The red curve indicates the ground-true contour.
	\end{sloppypar}
\end{figure}

\subsection{Limitations and future work}
Although comparison and ablation experiments on four ultrasound datasets have well proven the superiority of the BAAF module for medical ultrasound segmentation tasks. However, there are still some limitations of the study that need to be noted. From the ablation experiments, it can be seen that the introduction of the adaptive calibration mechanism (ACM) can further improve the segmentation performance of the network, but the mathematical evidence of the ACM's working mechanism is lacking. In addition, the prediction results illustrated in Fig. \textcolor{red}{7}, Fig. \textcolor{red}{9}, and Fig. \textcolor{red}{11} show that our method still needs to be optimized to reduce the occurrence of missed and false detections. Since the BAAF is an improvement of the existing attention mechanism, the introduction of the adaptive attention module will increase the network parameters and computational overhead inevitably. Therefore, there are several aspects that can be further considered in future research work: (1) enhancing the investigation of the interpretability and mathematical proof of the method; (2) refining the feature adaptive selection mechanism further (such as: intra-class adaptive and inter-class adaptive); and (3) Reducing network complexity while ensuring network performance.

\section{Conclusion}
To cope with the challenges arising from the ultrasound pattern complexity and morphology variety, we proposed a Basic Attention Adaptive Framework (BAAF) to assist doctors in automatic segmentation of medical ultrasound. The parallel hybrid attention module (PHAM) in BAAF can help the network to be calibrated coarsely from spatial and channel dimensions. Subsequently, the adaptive calibration mechanism (ACM) can further optimize the "what" and "where" of PHAM focus to capture more robust object characterizations. To verify the effectiveness of the proposed method, we have conducted extensive comparisons with state-of-the-art segmentation methods on four ultrasound datasets, and the experimental results fully illustrate the feasibility of our method. Overall, BAAF achieved satisfactory results in the medical ultrasound segmentation task, which provide a possibility for future clinically assisted diagnosis.

\bibliographystyle{IEEEtran}
\bibliography{IEEEabrv,tmi}

\end{document}